\documentclass[aps,prl,preprint,superscriptaddress]{revtex4}
\usepackage{graphicx,epsfig,pstcol,pstricks,amssymb}

\usepackage{hyperref}
\hypersetup{colorlinks=true,urlcolor=blue}
\begin{document}
{\small Molecular Physics, volume {\bf 103} pp. 1-5 (2005) \htmladdnormallink{http://dx.doi.org/10.1080/00268970412331293820}{http://dx.doi.org/10.1080/00268970412331293820}}  
\title{RESEARCH NOTE\\ The range of meta stability of ice-water melting for two simple models of water.}
\author{Carl McBride}
\author{Carlos Vega}
\author{Eduardo Sanz}
\author{Luis G. MacDowell}
\author{Jose L. F. Abascal}
\affiliation{
Departamento  de Qu\'{\i}mica F\'{\i}sica.
Facultad de Ciencias Qu\'{\i}micas. Universidad Complutense de Madrid.
Ciudad Universitaria 28040 Madrid, Spain.
}
\date{July 30, 2004}
\begin{abstract}
A number of crystal structures of water have been `superheated'
in Monte Carlo simulations. Two well known models for water
were considered; namely the TIP4P model and the SPC/E model.
By comparing the fluid-solid coexistence temperature to the
temperature at which the solid becomes mechanically unstable and
melts it is possible to determine the typical range of temperatures
over which is possible to superheat the ice phases in conventional
simulation studies. It is found that the ice phases can be superheated
to approximately  90K beyond the fluid-solid coexistence temperature.
Beyond this limit they spontaneously melt.
This limit  appears to depend weakly both on the type of ice phase
considered and on the chosen model.
Obviously only rigorous
free energy calculations can determine the equilibrium fluid-solid
coexistence of a model. However,  a ``rule of thumb" is that,
by subtracting 90K from   the mechanically stability
limit of the the ice phase one is provided with  a first guess as to  the equilibrium
fluid-solid coexistence temperature.
\end{abstract}

\maketitle
\section{Introduction}
Ever since the advent of statistical mechanical `experiments' on 
fast computing machines it was realized that performing computer 
simulations of water would be of the utmost importance to 
the understanding of what must be one of the most important molecules known to man.
Pioneering studies of such nature were performed by
Barker and Watts 
\cite{CPL_1969_3_0144_nolotengo} 
and by Rahman and Stillinger 
\cite{JCP_1971_55_03336_nolotengo}.
Since then, thousands of computer simulation studies have
been carried out. However,
the possibility of determining the water phase diagram by
computer simulation has not received such widespread attention. 
This is surprising since the phase diagram for a number
of molecular models such as spherocylinders
\cite{JCP_1996_104_06755,JCP_1997_106_00666}, 
linear tangent hard spheres 
\cite{MP_1995_85_0193_photocopy,JCP_2001_115_04203} 
and Gay Berne 
\cite{JCP_2002_117_06313,JCP_1997_107_08654} 
models are well known.  Although several
studies have examined the vapor-liquid equilibria of 
different model potentials of water
\cite{JPCB_1998_102_01029}, 
the fluid-solid equilibria has been investigated in just a few cases
\cite{JCP_1995_103_09744_nolotengo}.
Recently we have determined the phase diagram of two of the most
popular model potentials of water  
\cite{PRL_2004_92_255701,JCP_2004_121_01165}, 
namely the SPC/E 
\cite{JPC_1987_91_06269_nolotengo}
and the TIP4P 
\cite{JCP_1983_79_00926_nolotengo}
models (note that for a theoretical description, simpler models
may be required, based either on associating site potentials
\cite{JML_1997_7374_0317,MP_2001_99_0531,JPCB_1999_103_10272_nolotengo} 
or on polar convex bodies 
\cite{PCCP_2003_05_2391,MP_1992_76_0327_nolotengo}). 
In this way it was possible to show  that the simple
TIP4P model is able to provide a qualitatively correct view of the
phase diagram of water. In order to determine  the phase diagram, hundreds
of $NpT$ simulations were performed, leading to an equation of state for 
both the fluid and solid phases.
It was also necessary to compute the
free energy of the fluid phase (via thermodynamic integration) 
and the free energy of the solid phase (via Einstein crystal 
calculations 
\cite{JCP_1984_81_03188}). 
Once a single point on the
coexistence line was determined,
Gibbs-Duhem integration 
\cite{JCP_1993_98_04149} 
was used to obtain the full saturation line. 
Such calculations have allowed the authors to determine 
the phase diagram of the potential models TIP4P and  SPC/E 
and to establish their ability to reproduce the experimental
phase diagram of water
\cite{PRL_2004_92_255701}.
It is fair to say that the determination of the phase diagram
of a given  model potential of water is a cumbersome task. 
One may naively wonder as to whether $NpT$ runs could be sufficient
in order to obtain directly the fluid-solid equilibria of a simple
model. Unfortunately this is not possible.
When a ``molecular liquid" is cooled to below the freezing
temperature at constant pressure in an $NpT$ simulation, one 
usually obtains a
supercooled liquid. It is very difficult to  observe in computer
simulations the formation of a 
perfect crystal (also in experiments one often finds  
supercooled liquids).

What is the behavior of the solid
phase when heated at constant pressure?  Experimentally, when
a solid is heated at constant pressure it melts at the
melting temperature, because the surface acts as a nucleation
site. It is therefore not   possible to  superheat
a solid above the melting temperature. This sounds good since
it suggests a procedure to determine the melting temperature
from computer simulations; one simply heats the solid until it melts.
However, in practice this is not the case.
In computer simulations (in contrast to real experiments)
one may superheat the solid before it melts. 
This is well known for hard spheres 
\cite{allen_book} 
(with pressure being the thermodynamic variable in question)
and  for 
Lennard-Jones (LJ) particles 
\cite{DP_2002_47_00667_nolotengo,JCP_2004_120_11640}. 
In $NpT$ runs it is found that
the solid melts at pressures below the equilibrium melting
pressure (for hard spheres), or at temperatures above the
melting temperature (for the Lennard-Jones system).

Since the rigorous
phase diagram of water of two simple models is now available, it is possible,
for the first time, to analyze the typical range of temperatures
over which the  solid phases of water (ices) can be superheated
in a computer 
simulation before spontaneous melting occurs. 
The probability of melting once the ice is superheated obviously depends
on the size of the system and on the length of the run.
However, here our intention is to  provide `ball-park'  figures of the stability
range of the ice phases.
The numbers obtained may prove to be useful when designing 
new potential models which lead to a  better description of the
phase diagram of water.  
\section{Simulation Details}
The initial solid configurations were constructed 
using crystallographic data (taken from Ref.
\cite{bookPhysIce}
and references therein). 
In the case of the proton ordered ices 
(i.e. II and VIII).
this is all that is required.
However, for the proton disordered ices (i.e. I, VI and VII),
while the oxygens were situated on the lattice points, the hydrogen
atoms were located in disordered configurations such that the  
net dipole moment was zero as well as at the same time 
satisfying the ice rules 
\cite{JACS_1935_57_02680_nolotengo}.
This was done by using the algorithm of Buch et al.
\cite{JPCB_1998_102_08641_nolotengo}.
For ices III and V, which present a certain degree of proton
ordering
the Buch algorithm
was generalized in order to produce initial
configurations having biased occupation of the
hydrogen positions.

Anisotropic $NpT$ Monte Carlo
simulations (Rahman--Parrinello like)  were used for
the solid phases 
\cite{frenkelbook}.
The pair potential was truncated for all phases  at $8.5$ \AA.
Standard long range corrections to the LJ energy were added.
Ewald sums were employed for electrostatic interactions.

The number of particles used in the simulations is presented
in Table I (chosen for each solid phase so as
to allow for at least twice the cutoff distance in each direction).

The melting transition is monitored by following the progress of the
structure factor of the system.
The structure factor for the Bragg reflection of the planes $hkl$ of
the crystal is given by:
\begin{equation}
F_{hkl} = \frac{1}{N} \sum_{i=1}^{i=N} f_i \exp \left( 2 \pi i ( hx_i + ky_i + lz_i)\right)
\label{sok}
\end{equation}
The intensity of a given line is given by
\begin{equation}
I_{hkl} = |F_{hkl}|^2 = F_{hkl} F_{hkl}^*
\label{sokI}
\end{equation}

It should be  mentioned that only oxygens were used when computing
the structure factor in equation \ref{sok}.  The factor $f_{i}$ of oxygen
was arbitrarily set to one. For each solid structure
(ice I, II, III, V, VI)  the combination of $hkl$ values
that provided the most intense line were used to detect the melting transition.

The runs were performed by taking an initial crystalline configuration
under thermodynamic conditions corresponding to that of the solid phase.
This initial state was then simulated in intervals of 10 K
with runs of $8 \times 10^4$ cycles per temperature.
One cycle is defined as a trial
move per particle (translation or rotation) plus a trial volume
change.  
Each subsequent simulation was started from the final configuration
of the previous run.
When the structure factor was seen to fall to zero then the previous
temperature was re-run up to three times in order
to see whether this state too would
melt.
\section{Results}
A typical fall in the 
structure factor of an ice phase is shown 
in figure 1 (in this case for the melting of TIP4P--ice V  at
$T=$310 K and 0.5 GPa).
As can be seen, once the structure factor falls below
a certain value  the melting proceeds rapidly and irreversibly.
Results for the other ice phases  and models are similar.

Table \ref{superheat_tab} presents the first temperature for which  spontaneous melting
of ices was found (see the column labeled as
$T_{stab}$).
It should be noted that in all cases, simulation runs
of up to $2.4 \times 10^5$  cycles were performed for a lower temperature (by 5 or
10 K) without success in melting the ice.
In table \ref{superheat_tab}, the melting temperatures 
of the corresponding ice phase for the TIP4P and SPCE 
models are also shown for comparison 
(column denoted as $T_{coex}$). The latter values are taken from 
\cite{PRL_2004_92_255701} and 
were calculated by
determining the free energies of the fluid and solid phases.
$\Delta T$ (we shall denote this
value as the meta stability range) represents the difference
between $T_{stab}$ and $T_{coex}$ and is also
given in Table II.
For ice I the stability range depends only weakly  on
pressure (for the two pressures considered the range is   about
$\Delta T=68$K).
The stability range of different ices of a certain model
are slightly different, although these differences are never large.
For the TIP4P, we may state that 80 K is the typical 
range of meta stability of the different ice phases. For the
SPC/E model the meta stability range 
appears to be about 90 K (i.e.,
10 K larger than that of the SPC/E).
This is not surprising since the internal energies of the
solids are always lower in the SPC/E model than in the
TIP4P model (the hydrogen bonding is slightly stronger
in the SPC/E model when compared with  the TIP4P model). In any
case, differences
in the stability range of both models (SPC/E and TIP4P)
are small. 
In this respect, the numbers presented
here represent the typical values of the stability range that
would be expected to be found in simulations of other realistic models of water.
Although a systematic study of the system size dependence of $T_{stab}$
has not been performed, we have studied the behavior of $T_{stab}$ for
ice Ih of the TIP4P model at $p=0.1$ MPa for two system 
sizes: 288 molecules and 432 molecules.
The $50\%$ increase in system size delays the onset of melting by 
about 10 K. 

Finally, figures 2 and 3 show the location of the stability temperature
(symbols) of the different ices  for
both the TIP4P (figure 2) and
for the SPC/E (figure 3). The phase diagram as obtained from
free energy calculations is also shown (lines).
It can be seen that the degree of superheating is of the order of
85 K. Interestingly the stability temperatures reflects the
trends found in the equilibrium coexistence lines. This
suggests that a first rough estimate of the melting temperature of ices
can be obtained from the $NpT$ simulations.
Although for the SPC/E model ice III and V (and also ice Ih!) 
are not thermodynamically stable phases (i.e for any given $p$ and $T$ another
ice always exists with lower Gibbs free energy) they are 
mechanically stable and it is possible to perform simulations 
of these phases
\cite{PRL_2004_92_255701}.
The stability limit has been studied only for
the thermodynamically stable phases of the TIP4P and SPC/E models.
However it is also possible to determine $T_{stab}$ for ices
which are metastable with respect to other solid structures.
 For example for ice III in the SPC/E model at $p=$ 0.5 GPa it was found that 
$T_{stab}=270$ K, which is substantially lower 
than the value obtained for ice II (the thermodynamically
stable phase of the SPC/E at this pressure) at the same pressure,
having $T_{stab}=365$ K. 

\section{Acknowledgments}
\begin{acknowledgments}
Financial support from project numbers FIS2004-06227-C02-02 and
BFM-2001-1017-C03-02 of
the MCYT (Ministerio de Ciencia y Tecnolog\'{\i}a) 
is acknowledged.
C.M., would like to thank the Comunidad de Madrid 
for the award of a 
post-doctoral research grant (part funded by the European Social Fund). E.S. would like to thank MEC for
a predoctoral grant.  L.G. MacDowell would like to thank 
the Universidad Complutense de Madrid and MCYT for 
the award of a Ram\'on y Cajal fellowship.
\end{acknowledgments}
\newpage
\bibliography{bibliography}

\newpage
Captions to the figures:
~\\
~\\
Figure 1: A plot of the intensity of the structure factor,
$I_{hkl}$, for the TIP4P model of ice V at 310 $K$ and 0.5GPa.
This plot is representative of the sudden fall in the intensity of the structure factor associated with melting.
~\\
~\\
Figure2: Plot of the phase diagram of water for the TIP4P model.
Points are the temperatures at which the solid melted under constant pressure.
$\blacksquare$ ice I (288 molecules), $\square$  ice I (432 molecules), $\blacktriangledown$ ice III,  
$\blacklozenge$ ice V,
and $\blacktriangle$ ice VI.
~\\
~\\
Figure 3: Plot of the phase diagram of water for the SPC/E model.
Points are the temperatures at which the solid melted under constant pressure.
$\blacksquare$ ice I, $\bullet$ ice II, and $\blacktriangle$ ice VI.
\newpage
\begin{table}
 \begin{tabular}{lr}
\hline
Ice  &    N$^{\rm o}$. of molecules   \\
\hline
I(Ih) & 288\\
I(Ih) & 432\\
II & 432\\
III &324\\ 
V & 504\\
VI & 360\\
\hline
\label{nmolec_tab}
\end{tabular}
\caption{Relation between the number of molecules in the simulation box and
the ice structure simulated for both the SPC/E and the TIP4P models.
}
\end{table}
~\\
~\\
~\\
~\\
~\\
~\\
~\\
~\\
~\\
~\\
~\\
~\\
~\\

\newpage
\begin{table}
 \begin{tabular}{lrrrrrrr}
\hline
 &  &  TIP4P  &  &  & SPC/E &  \\
Ice & P (GPa) &  $T_{\rm coex}$   & $T_{\rm stab}$ & $\Delta T$ & $T_{\rm coex}$ &$T_{\rm stab}$ & $\Delta T$\\
\hline
I (288)   & 0.0001      & 232 & 300 &  68 &  215 & 295 & 80   \\
I (432)   & 0.0001      & 232 & 310 &  78 & ---  & --- & ---  \\
I    & 0.05    & 229 & 297 &  68 &   ---& --- &---   \\
II    & 0.5   &---     & --- &  ---& 249  &365  & 116  \\
III   & 0.3   & 197  & 280 &  83 & ---  &---  &---   \\
V     & 0.5   & 204  & 310 & 106 &  --- & --- & ---  \\
VI    & 2.0  & 270    & 350 &  80 &  --- &---  &---   \\
VI    & 2.5  & ---    &---  & --- &  234 & 330 & 96   \\
\hline
\end{tabular}
\caption{Results for the range of superheating for the various ices and
models. $T_{stab}$ is the temperature at which the system
becomes unstable (spontaneous melting), while $T_{coex}$ is the
thermodynamic melting temperature.
}
\label{superheat_tab}
\end{table}

\newpage
\begin{figure}
\includegraphics[clip]{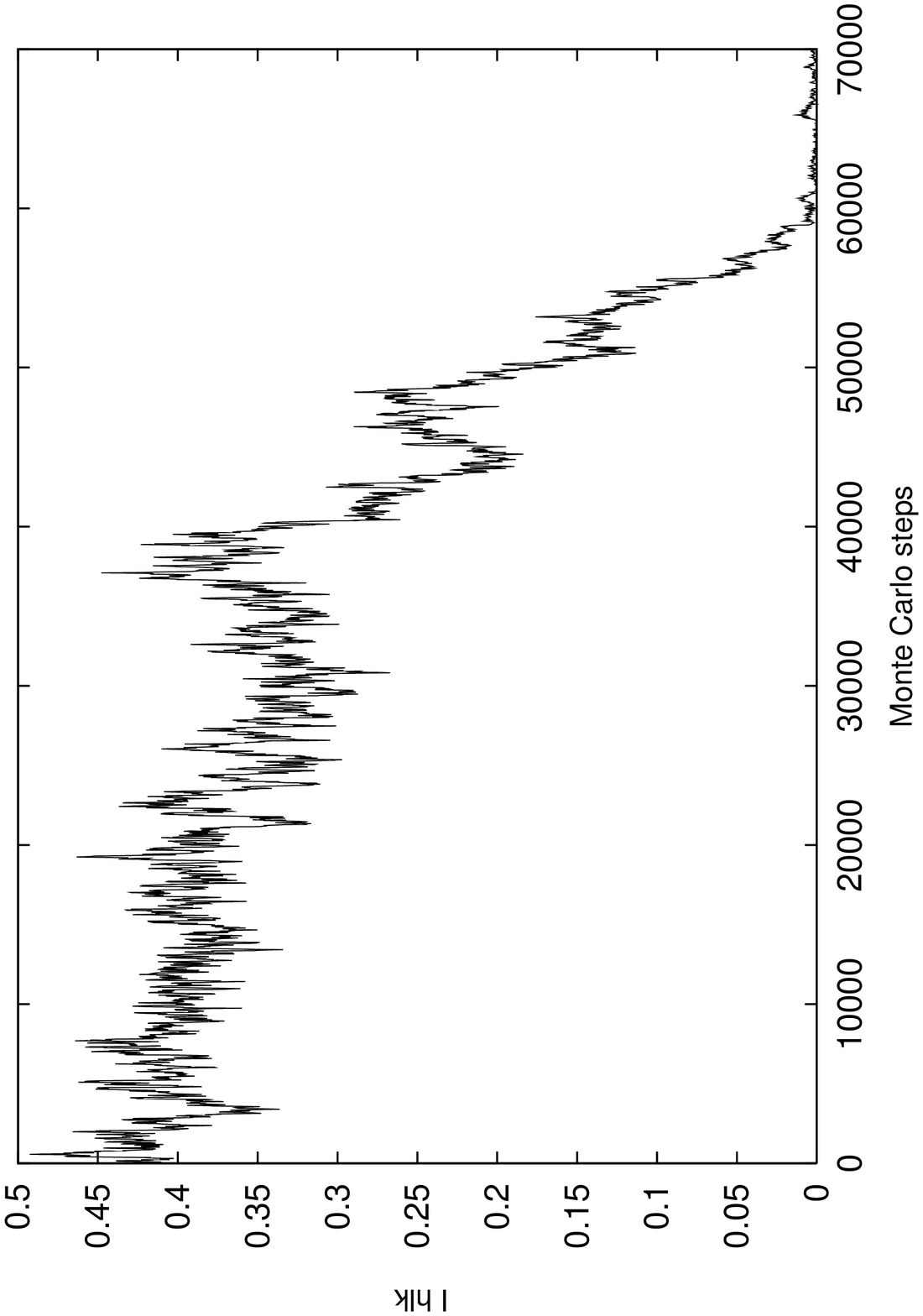}
\label{decay_fig}
\end{figure}

\newpage
\begin{figure}
\includegraphics[clip]{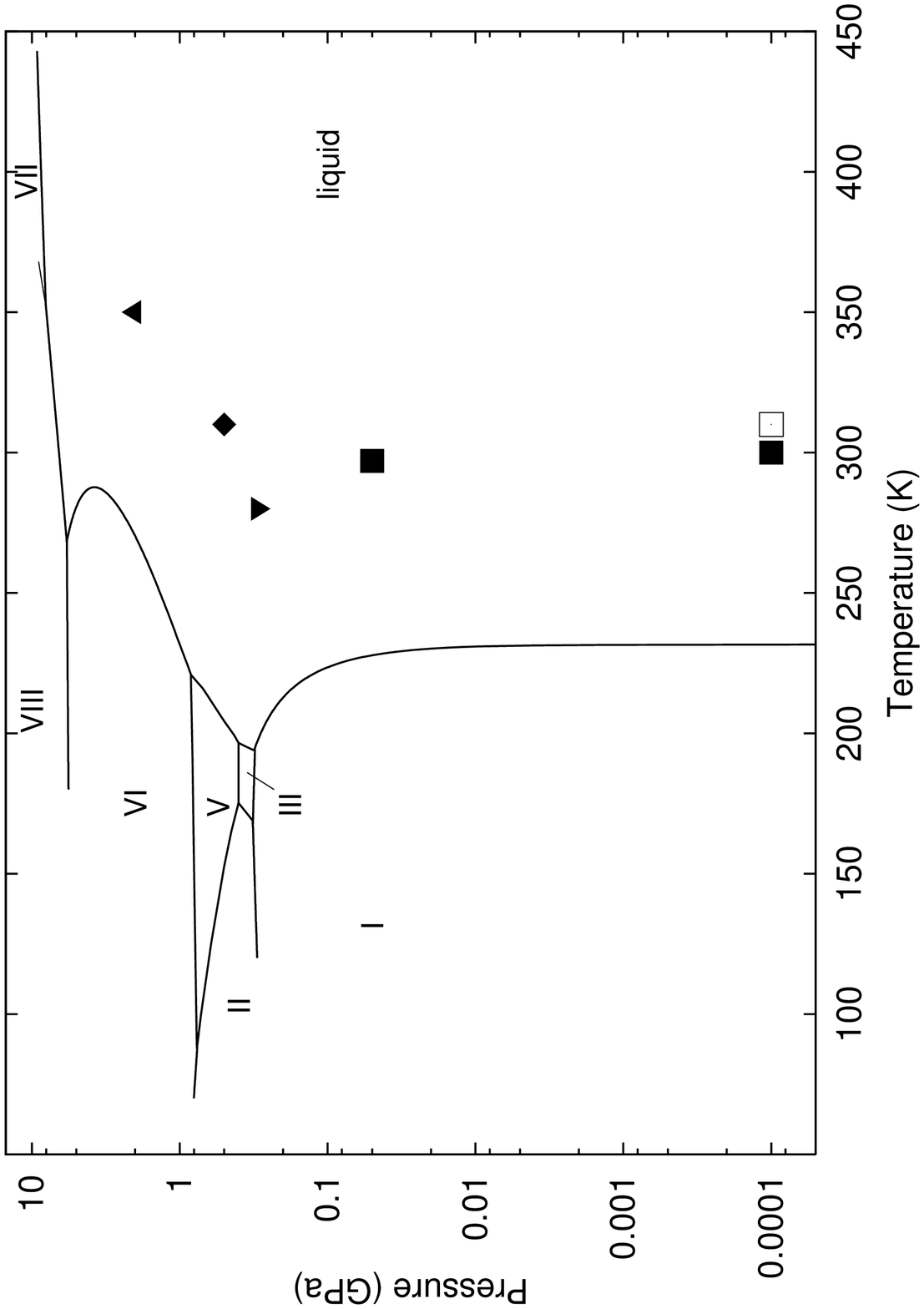}
\label{TIP_fig}
\end{figure}

\newpage
\begin{figure}
\includegraphics[clip]{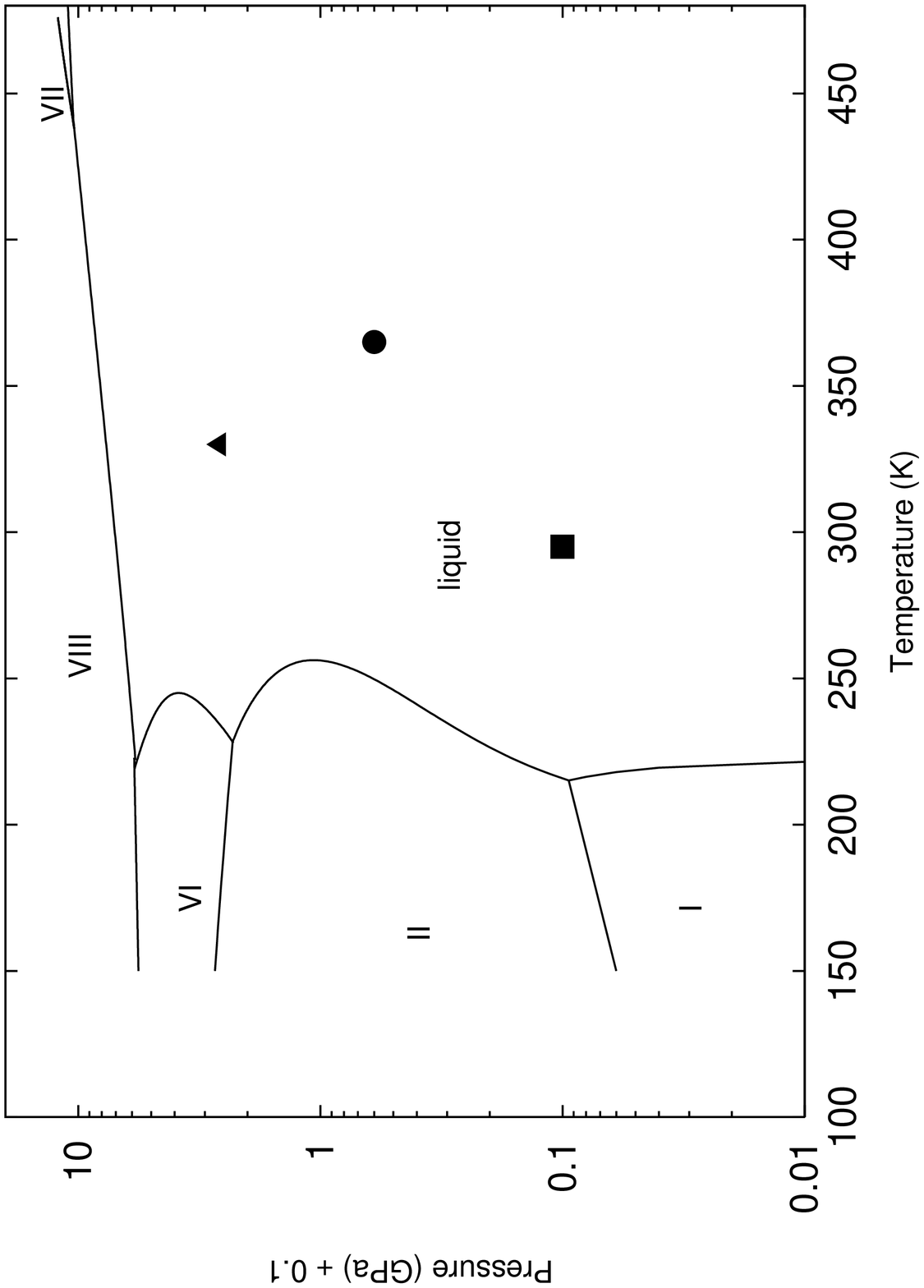}
\label{SPC_fig}
\end{figure}

\end{document}